\documentstyle[12pt]{article}

\begin{document}
\newcommand{\pl}[1]{Phys.\ Lett.\ {\bf #1}\ }
\newcommand{\npb}[1]{Nucl.\ Phys.\ {\bf B#1}\ }
\newcommand{\prd}[1]{Phys.\ Rev.\ {\bf D#1}\ }
\newcommand{\prl}[1]{Phys.\ Rev.\ Lett.\ {\bf #1}\ }

% draw box with width #1pt and line thickness #2pt
\newcommand{\drawsquare}[2]{\hbox{%
\rule{#2pt}{#1pt}\hskip-#2pt%  left vertical
\rule{#1pt}{#2pt}\hskip-#1pt%  lower horizontal
\rule[#1pt]{#1pt}{#2pt}}\rule[#1pt]{#2pt}{#2pt}\hskip-#2pt%  upper
\rule{#2pt}{#1pt}}% right vertical

% Young tableaux
\newcommand{\Yfund}{\raisebox{-.5pt}{\drawsquare{6.5}{0.4}}}%  fund
\newcommand{\Ysymm}{\raisebox{-.5pt}{\drawsquare{6.5}{0.4}}\hskip-0.4pt%
        \raisebox{-.5pt}{\drawsquare{6.5}{0.4}}}%  symmetric second rank
\newcommand{\Yasymm}{\raisebox{-3.5pt}{\drawsquare{6.5}{0.4}}\hskip-6.9pt%
        \raisebox{3pt}{\drawsquare{6.5}{0.4}}}%  antisymmetric second rank
\newcommand{\Ythree}{\raisebox{-3.5pt}{\drawsquare{6.5}{0.4}}\hskip-6.9pt%
        \raisebox{3pt}{\drawsquare{6.5}{0.4}}\hskip-6.9pt
        \raisebox{9.5pt}{\drawsquare{6.5}{0.4}}}

% Journals
\newcommand{\jref}[4]{{\it #1} {\bf #2}, #3 (#4)}
\newcommand{\NPB}[3]{\jref{Nucl.\ Phys.}{B#1}{#2}{#3}}
\newcommand{\PLB}[3]{\jref{Phys.\ Lett.}{#1B}{#2}{#3}}
\newcommand{\PR}[3]{\jref{Phys.\ Rep.}{#1}{#2}{#3}}
\newcommand{\PRD}[3]{\jref{Phys.\ Rev.}{D#1}{#2}{#3}}
\newcommand{\PRL}[3]{\jref{Phys.\ Rev.\ Lett.}{#1}{#2}{#3}}
\newcommand{\PRV}[3]{\jref{Phys.\ Rev.}{#1}{#2}{#3}}

\renewcommand{\thefigure}{\arabic{figure}}
\setcounter{figure}{0}
\begin{titlepage}
\begin{center}
{\hbox to\hsize{hep-th/9701191 \hfill  MIT-CTP-2605}}
{\hbox to\hsize{               \hfill  BU/HEP-97-3}}
{\hbox to\hsize{               \hfill  UCB-PTH-97/04}}
\bigskip

\bigskip

{\Large \bf  Self-Dual $N=1$ SUSY Gauge Theories }

\bigskip

\bigskip

{\bf Csaba Cs\'aki$^{a,}$\footnote{Address after August 1: Department of
Physics,
University of California, Berkeley, CA 94720.},
Martin Schmaltz$^b$,
Witold Skiba$^{a,}$\footnote{Address after September 1: Department of  
Physics,
University of California at San Diego, La Jolla  CA 92093.}
and \\
John Terning$^c$}\\

\smallskip

{ \small \it $^a$Center for Theoretical Physics,
Massachusetts Institute of Technology, Cambridge, MA 02139, USA }

\smallskip

{\tt csaki@mit.edu, skiba@mit.edu}

\bigskip

{\small \it $^b$Department of Physics, Boston University,
Boston, MA 02215, USA }

\smallskip

{\tt schmaltz@abel.bu.edu}

\bigskip

{\small \it $^c$ Department of Physics, University of California, Berkeley,
CA 94720, USA }

\smallskip

{\tt terning@alvin.lbl.gov}

\bigskip

\vspace*{1cm}
{\bf Abstract}\\
\end{center}
We present a large set of new self-dual $N=1$ SUSY gauge 
theories. Examples include $SU(N)$ theories with
tensors and $SO(N)$ theories with spinors.
Using these dualities as starting points, new non-trivial duals can be
derived by higgsing the gauge group or by integrating out matter.
General lessons that can be learned from these duals are:
``accidental" infrared symmetries play an important role in duality,
many theories have more than one ``dual", and
there seems to be no simple organizing pattern which relates
duals of theories with different number of flavors.

\bigskip

\bigskip

\end{titlepage}

\section{Introduction}

Following the pioneering work by Seiberg on duality for
supersymmetric QCD (SQCD)~\cite{Seib}, a number of examples of duality in
non-Abelian $N=1$ supersymmetric gauge theories have been
found~[2-13]. The well known examples are for theories with matter in
the fundamental representation of the gauge
group~[1-4], but there
are also examples for theories with spinor
representations~\cite{PouliotStrassler}.

Generalizing these dualities to theories with more general
matter representations has proven to be very difficult. For theories
with a tensor representation, much progress
has been made by adding suitable superpotential terms which
give a mass to the tensor away from the origin of moduli
space~\cite{Kutasov}.
Then out in moduli space the theories reduce to theories with
fundamentals only, to which dual descriptions are known.
Continuing these duals to the origin of moduli space, duals for
the theory with the tensor and the superpotential terms were
found. One might be tempted to try to obtain the theory without
a superpotential by taking the coupling of the superpotential
term to zero. However, it is clear that this limit is singular, since
now there are components of the tensor which become massless
even on generic points of moduli space. A weakly coupled
low energy description which includes all massless modes still
has not been found.

In this letter we propose a number of new dualities for $SU$ theories
with tensor matter and $SO$ theories with spinors, both with
no tree-level superpotential. Our proposed duals have the
interesting feature that they are self-duals. The electric theory and
its magnetic dual have identical gauge groups and gauge variant
matter fields. In addition to the dual gauge degrees of freedom
the magnetic theories also contain some number of fundamental
gauge invariant ``meson" fields which are coupled to the gauge variant
fields in the superpotential.

Self-duals have been known to exist for theories with matter
only in fundamental
representations~\cite{Seib,IntSeib,IntPouliot,Ramond,Distler}.
For example, consider
SQCD with $N$ colors and the special number of
flavors $F=2N$. The field content and symmetry properties of this
theory are summarized in the following table.

\begin{equation}
\begin{array}{c|c|cccc}
& SU(N) & SU(2N) & SU(2N) & U(1) & U(1)_R \\
\hline
Q & \Yfund  & \Yfund & 1  & 1 & 1/2\\
\bar{Q} & \overline{\Yfund} & 1 & \Yfund & -1 & 1/2\\
\end{array}
\end{equation}

The dual for this theory is just a special case of Seiberg's dual
for SQCD. It is also an $SU(N)$ gauge theory with dual quarks
$q$ and $\overline{q}$ and a fundamental ``meson" gauge singlet
field $M$ coupled in the superpotential with the term $W=M q \bar{q}$.

\begin{equation}
\begin{array}{c|c|cccc}
& SU(N) & SU(2N) & SU(2N) & U(1) & U(1)_R \\
\hline
q & \Yfund  & \overline{\Yfund} & 1  & 1 & 1/2\\
\bar{q} & \overline{\Yfund} & 1 & \overline{\Yfund} & -1 & 1/2\\
M & 1  & \Yfund & \Yfund  & 0 & 1\\
\end{array}
\end{equation}

Note that the standard consistency checks are rather trivial for this
self-dual. For example, the
anomaly matching conditions are almost all trivially satisfied. All
anomalies involving $U(1)$ and $U(1)_R$ charges are
matched because the fermion component of $M$ is uncharged under
the $U(1)$'s and the contributions of the dual quarks are identical
to the contributions of the electric quarks. The only non-trivially
matched anomalies are the $SU( 2N)^3$ non-Abelian flavor anomalies.
The operator maps of the gauge invariant ``mesons" and ``baryons"
are also very simple:

\begin{eqnarray}
Q\bar{Q}&\leftrightarrow& M  \nonumber \\
 Q^N &\leftrightarrow&  q^N \nonumber \\
\bar{Q}^N &\leftrightarrow&  \bar{q}^N
\end{eqnarray}

The dual theory has an additional gauge invariant $q\bar{q}$
which is set to zero by the equations of motion derived from the
superpotential. Since these operators parameterize the moduli spaces of the
two theories, this operator map is necessary to show the equivalence
of the moduli spaces. Note that one can obtain the dualities for all other
numbers of flavors by integrating out flavors from one of the two theories.

Similar self-duals exist for theories with ``quarks" transforming as
fundamentals of $Sp$~\cite{IntPouliot}, $SO$~\cite{IntSeib},
and some exceptional gauge groups~\cite{Ramond,Distler}.
Theories with matter in tensor representations and certain
simplifying tree-level superpotentials have also been found to
have self-duals~\cite{Kutasov}. More recently,
a self-dual has been proposed for a theory with a tensor field
and no tree-level superpotential~\cite{oursp}. Here, we
present similar duals for a number of theories with matter in tensor
and spinor representations.

In the following section we review Seiberg's proposed dual of
$SU(2)$ SQCD, which requires the existence of accidental
infrared symmetries. We then describe three different duals of
this theory.
In the third section we list a number of new self-duals and other
duals which can be derived via flows from the self-duals. For some
of the presented dualities we also give detailed consistency checks.
While the number of self-duals that we present is quite large, it is
clear that this list is not exhaustive and it should be possible
to generate more examples with similar properties. We present our
conclusions in the final section.

\section{Multiple self-duals and accidental symmetries}

In this section we review two apparently common but not very well known
features of duality. One is the occurrence of accidental infrared
symmetries~\cite{Distler,LeighStrassler}, and the other is the existence
of multiple ``duals" for a single theory~\cite{LST,otherduality}.

Frequently, the ultraviolet description of one (or several) of these duals
does not manifestly have the full global symmetry of the infrared.
Only in the far infrared, below the intrinsic scale of the strong
gauge interactions, does the full global symmetry become
manifest. Let us illustrate this with the example of
$SU(2)$ SQCD for which the fundamental and antifundamental
representations are equivalent. Thus, the full non-Abelian flavor
symmetry is $SU(2F)$ rather than just $SU(F) \times SU(F) \times U(1)$,
and the spectrum of the theory falls into $SU(2F)$
representations for all numbers of flavors $F$.
The independent gauge invariant chiral operators which parameterize the
moduli space and correspond to infrared degrees of freedom are contained
in the matrix $A=QQ$ which transforms in the antisymmetric two index tensor
representation of $SU(2F)$.

In Seiberg's dual description, discussed in the previous section,
the full flavor symmetry is not manifest, and one might be worried that
the duality is not valid for $SU(2)$. Seiberg's proposed dual and its
ultraviolet-symmetry for arbitrary $F$ are

% table of the dual of SU(2) SQCD
\begin{equation}
\begin{array}{c|c|cccc}
& SU(F-2) & SU(F) & SU(F) & U(1) & U(1)_R \\
\hline
q & \Yfund  & \overline{\Yfund} & 1  & 2/(F-2) & 2/F \\
\bar{q} & \overline{\Yfund} & 1 & \overline{\Yfund} & -2/(F-2) & 2/F \\
M & 1  & \Yfund & \Yfund  & 0 & 2-4/F \\
\end{array}
\end{equation}

Thus the ultraviolet global symmetries of this description contain
only $SU(F) \times SU(F) \times U(1) \subset SU(2F)$.
In order for the infrared spectrum to respect the full $SU(2F)$
flavor symmetry the ``meson" $M$ has to be complemented
by bound states of the dynamical dual quarks $q$ and $\bar{q}$. Together
these degrees of freedom transform as a full $SU(2F)$ representation.
Thus, the $SU(2F)$ symmetry generators mix fundamental fields
with composites. In order for this to make any sense, the
fields which are to form a complete irreducible $SU(2F)$
representation have to have identical quantum numbers under
the other symmetries. Here, these extra quantum numbers are the
$U(1)_R$ charges. The chiral gauge invariants of the dual with
their global transformation properties are

% global charges of the dual of SU(2) SQCD
\begin{equation}
\begin{array}{c|cccc}
 & SU(F) & SU(F) & U(1) & U(1)_R \\
\hline
M & \Yfund & \Yfund  & 0 & 2-4/F \\
b=q^{F-2} & \Yasymm  & 1  & 2 & 2-4/F \\
\bar{b}=\bar{q}^{F-2} & 1 & \Yasymm & -2 & 2-4/F \\
\end{array}
\end{equation}

The composite operator $q\bar{q}$
is set to zero by the $M$ equation of motion. Note that the $U(1)_R$
charges of the all the non-vanishing chiral operators are identical, thus
it is possible to unify $M$, $b$, and $\bar{b}$ into the antisymmetric
tensor representation of $SU(2F)$.

Another necessary condition for the emergence of the full
$SU(2F)$ global symmetry is that the scaling dimensions of the three
operators $M$, $b$ and $\bar{b}$ have to agree.
In the range of $F$ where this theory has an
infrared fixed-point, the theory is superconformal, and  the scaling
dimensions are given by $3/2$ times the superconformal $R$-charges of
the fields. Since the $R$-charges commute with $SU(2F)$, the dimensions
respect the full flavor symmetry as well.
Note that the $SU(2)$ theory and its dual can be reached via flows
from the more general $SU(N)$ theory and its $SU(F-N)$ dual,
thus all the usual consistency checks also apply to this dual.

There is an additional dual of the $SU(2)$ theory which has a manifest
$SU(2F)$ symmetry. The field content of this $Sp$ dual is

% table for Sp dual here
\begin{equation}
\begin{array}{c|c|cc}
& Sp(2F-6) & SU(2F) & U(1)_R \\
\hline
q & \Yfund  & \overline{\Yfund} & 2/F \\
A & 1  & \Yasymm & 2-4/F \\
\end{array}
\end{equation}
In this dual the whole antisymmetric tensor $A=QQ$ is fundamental.

An interesting special case is the theory with $F=4$. This theory is
self-dual,
and both the $SU$-dual and the $Sp$-dual have an $SU(2)$ gauge group.
The two duals differ in the ultraviolet by their ``meson" content. While the
$Sp$ dual has a ``meson" $A=QQ$ which transforms as an antisymmetric tensor  
of the
full $SU(8)$ flavor symmetry, the $SU$ dual only has an $SU(4) \times SU(4)$
symmetry and the ``meson" transforms as $M=(\Yfund, \Yfund)$. Again, in the
infrared, the full flavor symmetry is restored and the ``meson" $M$ is  
unified into an $SU(8)$ multiplet together with the composites $b=qq$ and
$\bar{b}=\bar{q}\bar{q}$.
There is a third possible dual which also has an $SU(2)$ gauge symmetry,
and where the only fundamental ``meson" fields are the
``baryons" $B = (\Yasymm,1)$ and $\bar{B} = (1,\Yasymm)$.

We summarize the field content of the self-dual theory with its three duals  
in the following table.

%table with the three duals for N=2 and F=4
\begin{equation}
\begin{array}{c|c|cccc}
& SU(2) & SU(4) & SU(4) & U(1) & U(1)_R \\ \hline
Q & \Yfund  & \Yfund & 1 & 1 & 1/2 \\
\bar{Q} & \Yfund & 1 & \Yfund & -1 & 1/2 \\ \hline \hline
& SU(2)_{D1} & SU(4) & SU(4) & U(1) & U(1)_R \\ \hline
q & \Yfund  & \overline{\Yfund} & 1  & 1 & 1/2 \\
\bar{q} & \Yfund & 1 & \overline{\Yfund} & -1 & 1/2 \\
M & 1  & \Yfund & \Yfund  & 0 & 1 \\ \hline \hline
& SU(2)_{D2} & SU(4) & SU(4) & U(1) & U(1)_R \\ \hline
q & \Yfund  & \overline{\Yfund} & 1  & -1 & 1/2 \\
\bar{q} & \Yfund & 1 & \overline{\Yfund} & 1 & 1/2 \\
M & 1  & \Yfund & \Yfund  & 0 & 1 \\
B & 1  & \Yasymm & 1  & 2 & 1 \\
\bar{B} & 1 & 1 & \Yasymm & -2 & 1 \\ \hline \hline
& SU(2)_{D3} & SU(4) & SU(4) & U(1) & U(1)_R \\ \hline
q & \Yfund  & \Yfund & 1  & -1 & 1/2 \\
\bar{q} & \Yfund & 1 & \Yfund & 1 & 1/2 \\
B & 1  & \Yasymm & 1  & 2 & 1 \\
\bar{B} & 1 & 1 & \Yasymm & -2 & 1 \\
\end{array}
\end{equation}

The original theory and dual number 2 in this table have a manifest
$SU(2F)$ flavor symmetry. For purpose of comparison between the three duals
we listed only the transformation properties under the
$SU(F) \times SU(F) \times U(1)$ subgroup.
Note that the charges and representations of the dual quarks $q$ and
$\overline{q}$ are different in the three duals. Also, these duals
have differing superpotentials
\begin{eqnarray}
W_{D_1} &=& M q\bar{q} \nonumber \\
W_{D_2} &=& M q\bar{q} + B q^2 +\bar{B} \bar{q}^2  \\
W_{D_3} &=& B q^2 +\bar{B}\bar{q}^2 \ .\nonumber
\end{eqnarray}

As we will demonstrate in the following section, these phenomena are
quite common for self-dual theories. We will present several new examples of
self-dual theories with more than one dual, and duals which only have
a subset of the full flavor symmetry manifest in the ultraviolet.

%%%%%%%%%%%%%%%%%%%%%%%%%%%%%%
%%%%%%%%%%%%%%%%%%%%%%%%%%%%%%
\section{New examples of self-duals}

\subsection{$SU(8)$ with two antisymmetric tensors and 8 antifundamentals}

The first new example of self-dual theories that we present is based
on an $SU(8)$ gauge group with matter content $2\, \Yasymm +
8 \,\overline{\Yfund}$. The field content and the global symmetries of this
theory
and its dual is summarized in the table below.
\begin{equation}
\begin{array}{c|c|cccc}
& SU(8) & SU(2) & SU(8) & U(1) & U(1)_R \\ \hline
A & \Yasymm & \Yfund & 1 & 2 & 0 \\
\bar{Q} & \overline{\Yfund} & 1 & \Yfund & -3 & \frac{1}{2} \\
\hline \hline
& SU(8) & SU(2) & SU(8) & U(1) & U(1)_R \\ \hline
a & \Yasymm & \Yfund & 1 & 2 & 0 \\
\bar{q} & \overline{\Yfund} & 1 & \overline{\Yfund} & -3 & \frac{1}{2} \\
H=(A\bar{Q}\bar{Q}) & 1 & \Yfund & \Yasymm & -4 & 1 \\
K=(A^5\bar{Q}\bar{Q}) & 1 & \Yfund & \Yasymm & 4 & 1 \end{array}
\end{equation}
The tree-level superpotential in
the dual theory is~\footnote{In this paper we absorb the numerical
coefficients of the different superpotential terms into field
redefinitions.}
\begin{equation}
\label{SU8tree}
W=H a^5\bar{q}\bar{q}+K a\bar{q}\bar{q}.
\end{equation}

The 't~Hooft anomaly matching conditions are almost trivially satisfied,
since
the extra gauge singlets do not contribute to any anomaly involving
$U(1)_R$ while the $U(1)$ charges of the gauge singlets
come in charge conjugate pairs.
Thus the only non-trivial anomaly that is left to check is $SU(8)^3$
which matches between the electric and magnetic descriptions.

The correspondence of flat directions of the electric and magnetic theories
is straightforward as well:
\begin{eqnarray}
&A^4 \leftrightarrow a^4 & \nonumber \\
& \bar{Q}^8 \leftrightarrow \bar{q}^8& \nonumber \\
& A\bar{Q}^2 \leftrightarrow H & \nonumber \\
& A^5\bar{Q}^2 \leftrightarrow K \nonumber
\end{eqnarray}
The additional gauge invariants of the magnetic theory, $a\bar{q}^2$
and $a^5\bar{q}^2$ are set to zero by the $H$ and $K$ equations of
motion.

Further consistency checks of this proposed duality include higgsing the
gauge group by going out in different directions on the moduli
space\footnote{
Since this $SU(8)$ theory is completely chiral, one can not add mass terms
for any of the operators.}. First we consider giving a vacuum expectation
value
(VEV) to one of the antisymmetric tensor fields
\begin{equation} A_1 = v\left( \begin{array}{cccc} i\sigma_2 & & & \\
& i\sigma_2 & & \\
& & i\sigma_2 & \\
& & & i\sigma_2 \end{array} \right).
\end{equation}
This breaks the gauge group to $Sp(8)$, and the uneaten fields transforming
under $Sp(8)$ are $\Yasymm \, +8\, \Yfund$, with no 
superpotential.
The operator map determines that in the dual this corresponds
to giving a VEV to
\begin{equation}
\label{a1}
a_1 = w\left( \begin{array}{cccc} i\sigma_2 & & & \\
& i\sigma_2 & & \\
& & i\sigma_2 & \\
& & & i\sigma_2 \end{array} \right),
\end{equation}
which breaks the gauge group to $Sp(8)$ as well. The particle content
of the magnetic description is
\begin{equation} \begin{array}{c|c|c}
& Sp(8) & SU(8) \\ \hline
a & \Yasymm & 1 \\
q & \Yfund & \overline{\Yfund} \\
M_0 \equiv H_1& 1 & \Yasymm \\
M_1 \equiv H_2& 1 & \Yasymm \\
M_2 \equiv K_1& 1 & \Yasymm \\
M_3 \equiv K_2& 1 & \Yasymm \end{array}, \end{equation}
where we have decomposed $H$ and $K$ under the explicitly broken $SU(2)$
global
symmetry. The superpotential of equation~\ref{SU8tree} becomes
\begin{equation}
W=M_0 qa^3q+M_1qa^2q+M_2qaq+M_3qq
\end{equation}
after substituting the VEV of $a_1$ from equation~\ref{a1}. Thus we recover
the
self-dual of $Sp(8)$ with $\Yasymm +8\, \Yfund$ of Ref.~\cite{oursp}.

Another possible way to higgs the electric theory is by giving a VEV to
$A_1\bar{Q}_7 \bar{Q}_8$. This breaks the gauge group to $SU(6)$ and the
remaining matter content transforming under $SU(6)$ is $2 \, \Yasymm +
2 \, \Yfund + 6 \, \overline{\Yfund}$. On the magnetic side this corresponds
to giving a VEV to the $(7,8)$ component of  $H_1$, which
explicitly breaks some of the
global symmetries, but not the $SU(8)$ gauge symmetry. Thus, we obtain a
non-trivial dual of $SU(6)$ with $2 \, \Yasymm + 2 \, \Yfund + 6 \,
\overline{\Yfund}$, where the dual description is in terms of an $SU(8)$
gauge group with $2\, \Yasymm + 8\, \overline{\Yfund} +$ ``mesons",
and a complicated tree-level superpotential. Similar non-trivial dualities
derived  from self-duals will be described in detail in the next section.

%%%%%%%%%%%%%%%%%%%%%%%%%%%%%%%%%%%%%%%%%%%%%%%%%%%%%%%%%%%%%%%%%%%%%%%%%%%%% 

\subsection{$SU(6)$ with a three index antisymmetric tensor}
In this section we present a self-dual description for an $SU(6)$ gauge
theory with a three-index tensor and six flavors, $\Ythree + 6(\Yfund
+\overline{\Yfund})$. Starting from this duality
we derive several other non-trivial dualities by either higgsing the
electric gauge group or integrating out flavors. We will present these
dualities in detail. An important lesson to be learned from this set of
dualities is that in general dual theories do not follow any simple pattern.
As we will see, the tensor representations of the electric and magnetic
descriptions can be very different. Furthermore, theories with identical  
gauge
group
and gauge degrees of freedom but with slightly differing ``meson" content  
and
superpotential are dual to very different electric theories.

The field content and symmetries of the electric and magnetic theories are
\begin{equation}
\label{su6}
\begin{array}{c|cccccc}
 & SU(6) & SU(6)_Q & SU(6)_{\bar{Q}} & U(1)_1 & U(1)_2 & U(1)_R \\ \hline
 A & \Ythree & 1 & 1 & 0 & -2 & 0 \\
 Q & \Yfund & \Yfund & 1 & 1 & 1 & \frac{1}{2} \\
 \bar{Q} & \overline{\Yfund} & 1 & \Yfund & -1 & 1 & \frac{1}{2}
    \\ \hline \hline
 & SU(6)_D & SU(6)_Q & SU(6)_{\bar{Q}} & U(1)_1 & U(1)_2 & U(1)_R \\ \hline
 a & \Ythree & 1 & 1 & 0 & -2 & 0 \\
 q & \Yfund & \overline{\Yfund} & 1 & 1 & 1 & \frac{1}{2} \\
 \bar{q} & \overline{\Yfund} & 1 & \overline{\Yfund} & -1 & 1 & \frac{1}{2}
\\
M_0& 1 & \Yfund & \Yfund & 0 & 2 & 1 \\
M_2& 1 & \Yfund & \Yfund & 0 & -2 & 1
\end{array}
\end{equation}
The dual theory has the following tree-level superpotential
\begin{equation}
W=M_0 qa^2\bar{q} + M_2q\bar{q}.
\end{equation}
The global anomalies in the two descriptions again match almost trivially.
Flat directions in the two theories are also in one-to-one correspondence.
The mapping of the flat directions is given in the list below
\begin{equation}
\begin{array}{cc}
  A Q^3 \leftrightarrow a q^3 & Q^6 \leftrightarrow q^6 \\
  A \bar{Q}^3 \leftrightarrow a \bar{q}^3 &
        \bar{Q}^6 \leftrightarrow \bar{q}^6 \\
  A^3 Q^3 \leftrightarrow a^3 q^3 & Q \bar{Q} \leftrightarrow M_0 \\
  A^3 \bar{Q}^3 \leftrightarrow a^3 \bar{q}^3 &
          Q A^2 \bar{Q} \leftrightarrow M_2  \\
  A^4 \leftrightarrow a^4 &
\end{array}
\end{equation}
The extra $ qa^2\bar{q}$ and $q\bar{q}$ directions in the magnetic theory
are lifted by the $M_0$ and $M_2$ equations of motion.

Starting from the above presented dual for $SU(6)$ with $\Ythree + 6
(\Yfund + \overline{\Yfund})$ one can obtain new dualities by integrating
out a flavor or higgsing $SU(6)$ to $SU(5)$ or $SU(4)$. First, we consider
integrating out one flavor. The electric theory will be
$SU(6)$ with $\Ythree + 5(\Yfund + \overline{\Yfund})$.
In the dual description the quark mass term maps onto the
superpotential term $m (M_0)_{66}$. Thus
\begin{equation}
W=M_0 qa^2\bar{q} + M_2 q\bar{q} + m (M_0)_{66}.
\end{equation}
The $M_0$ and $M_2$ equations of motion force a VEV for the fields
$(q_6)_6=(\bar{q}_6)_6=a_{123}=a_{145}=v$ and zero VEVs for all other
components.
This breaks the $SU(6)$ gauge group to $Sp(4)$.
The resulting pair of dual theories is given in the table below.
\begin{equation}
\label{susixfive}
\begin{array}{c|cccccc}
 & SU(6) & SU(5)_Q & SU(5)_{\bar{Q}} & U(1)_1 & U(1)_2 & U(1)_R \\ \hline
 A & \Ythree & 1 & 1 & 0 & -5 & -1 \\
 Q & \Yfund & \Yfund & 1 & 1 & 3 & 1 \\
\bar{Q} & \overline{\Yfund} & 1 & \Yfund & -1 & 3 & 1
  \\ \hline \hline
 & Sp(4) & SU(5)_Q & SU(5)_{\bar{Q}} & U(1)_1 & U(1)_2 & U(1)_R \\ \hline
 a & \Yasymm & 1 & 1 & 0 & -10 & -2 \\
 q & \Yfund & \overline{\Yfund} & 1 & 3/2 & 2 & 1 \\
 \bar{q} & \overline{\Yfund} & 1 & \overline{\Yfund} & -3/2 & 2 & 1 \\
M_0& 1 & \Yfund & \Yfund & 0 & 6 & 2 \\
M_2& 1 & \Yfund & \Yfund & 0 & -4 & 0 \end{array}
\end{equation}

The new superpotential for the magnetic $Sp(4)$ theory is
\begin{equation}
W=M_0 q a \bar{q} + M_2 q \bar{q}.
\end{equation}

As a consistency check on the presented dual description we integrate
out one more flavor in the $SU(6)$ theory with a $\Ythree +
5(\Yfund + \overline{\Yfund})$. This theory with four flavors
is known to be confining~\cite{s-confinement}, and we should
reproduce its spectrum and the confining
superpotential from the dual description.
The $SU(6)$ theory with four flavors has the following confined
infrared spectrum:
$M_0=Q\bar{Q}$,
$M_2=QA^2\bar{Q}$,
$B_1=AQ^3$,
$\bar{B}_1 =A \bar{Q}^3$,
$B_3=A^3 Q^3$,
$\bar{B}_3=A^3 \bar{Q}^3$ and
$T=A^4$.
The confining superpotential in terms of these fields is given by
\begin{eqnarray}
\label{su6conf}
  W_{F=4}&=&\frac{1}{\Lambda^{11}} \Big( M_0 B_1\bar{B}_1T +B_3\bar{B}_3 M_0
  + M_2^3M_0+TM_2M_0^3+ \nonumber \\
  && \bar{B}_1B_3M_2+B_1\bar{B}_3M_2 \Big),
\end{eqnarray}

In the dual $Sp(4)$ description,
the mass term forces non zero VEVs for $a$ and $q_5$, $\bar{q}_5$, which
completely breaks the $Sp(4)$ gauge group. The fields $(M_0)_{55},
(M_0)_{5i},(M_0)_{i5},(M_2)_{55},(M_0)_{5i},(M_2)_{i5}$
and two components of $q_i$, $\bar{q}_i$ as well as two uneaten singlet
components of $a$ get masses from the tree-level superpotential.
The remaining components of
$q_i$, $\bar{q}_i$ are identified with $B_1$, $\bar{B}_1$, $B_3$ and
$\bar{B}_3$, the remaining singlet with $T$, and the remaining components of
$M_0$ and $M_2$ with the corresponding $M_0$ and $M_2$ in the $SU(6)$  
theory.
The tree-level superpotential reproduces the terms $M_0 B_1\bar{B}_1T$,
$B_3\bar{B}_3 M_0$, $\bar{B}_1B_3M_2$ and $B_1\bar{B}_3M_2$ of the confining
superpotential of equation~\ref{su6conf}, while the missing $ M_2^3M_0$ and
$TM_2M_0^3$ terms are presumably generated by instanton effects in the
completely broken $Sp(4)$.

Using the dual pair of the table in Eq.~\ref{susixfive} one can obtain a dual  
description
for $Sp(4)$ with an antisymmetric tensor and ten fundamentals.
We add conjugate ``meson" fields $\overline{M}_0$ and $\overline{M}_2$
with mass terms $W=M_0 \overline{M}_0 + M_2 \overline{M}_2$
to both theories.
Integrating out these massive ``mesons" in the $Sp(4)$ theory will set the
superpotential to zero. In the $SU(6)$ theory these mass terms
correspond to non-trivial interaction terms. The resulting dual of the
$Sp(4)$ theory with an antisymmetric tensor and ten fundamentals is
given in the following table.
\begin{equation}
\begin{array}{c|cccccc}
 & Sp(4) & SU(5)_Q & SU(5)_{\bar{Q}} & U(1)_1 & U(1)_2 & U(1)_R \\ \hline
 a & \Yasymm & 1 & 1 & 0 & -10 & -2 \\
 q & \Yfund & \Yfund & 1 & 3/2 & 2 & 1 \\
 \bar{q} & \overline{\Yfund} & 1 & \Yfund & -3/2 & 2 & 1 \\ \hline \hline
 & SU(6) & SU(5)_Q & SU(5)_{\bar{Q}} & U(1)_1 & U(1)_2 & U(1)_R \\ \hline
 A & \Ythree & 1 & 1 & 0 & -5 & -1 \\
 Q & \Yfund & \overline{\Yfund} & 1 & 1 & 3 & 1 \\
\bar{Q} & \overline{\Yfund} & 1 & \overline{\Yfund} & -1 & 3 & 1 \\
\overline{M}_0& 1 & \Yfund & \Yfund & 0 & -6 & 0 \\
\overline{M}_2& 1 & \Yfund & \Yfund & 0 & 4 & 2
\end{array}
\end{equation}
The dual has the tree-level superpotential
\begin{equation}
W=\overline{M}_0 Q \bar{Q} + \overline{M}_2 Q A^2 \bar{Q}
\end{equation}
in the magnetic $SU(6)$ theory.
Note that the electric $Sp(4)$ has an $SU(10)$ flavor
symmetry, while na\"{\i}vely the magnetic $SU(6)$ theory has only  
$SU(5)\times
SU(5)\times U(1)$. This is another example of accidental symmetries
analogous to the $SU(F-2)$ dual of $SU(2)$ described in detail in Section 2.

Next, we obtain dual descriptions of $SU(5)$ with matter content
$\Yasymm + \overline{\Yasymm}+ F(\Yfund + \overline{\Yfund})$
with $F=4$ or $5$ by higgsing the $SU(6)$ theory with a
three-index tensor and six flavors, which was described above.
We give a VEV to one flavor breaking
$SU(6)$ to $SU(5)$. The three-index antisymmetric tensor decomposes
into $\Yasymm$ and $\overline{\Yasymm}$ and there are five $SU(5)$ flavors
that remain uneaten. There remain also $SU(5)$ singlets
which can be eliminated by adding conjugate singlets and terms
\begin{equation}
W=S_i (q_i \bar{q}_6) + \bar{S}_i (q_6 \bar{q}_i), \; \; i=1,\ldots,5,
\end{equation}
to the superpotential. This extra term makes the unwanted fields massive
after $q_6$ and $\bar{q}_6$ get a VEV. In the dual description these terms
correspond to mass terms $S_i (M_0)_{i6}$ and $\bar{S}_i (M_0)_{6i}$,
which after integrating out these fields set all superpotential terms
involving $(M_0)_{i6}$ and $(M_0)_{6i}$ to zero. The VEV of $ (M_0)_{66}$
explicitly breaks the $SU(6)\times SU(6)$ global symmetry to $SU(5)\times
SU(5)$.
There are three non-anomalous $U(1)$ symmetries preserved by the tree-level
superpotential and the VEV of $M_0$. Thus, the resulting dual pair is
\begin{equation}
\begin{array}{c|ccccccc}
 & SU(5) & SU(5)_Q & SU(5)_{\bar{Q}} & U(1)_1 & U(1)_2 & U(1)_3 & U(1)_R
  \\ \hline
 A & \Yasymm & 1 & 1 & 0 & -5 & -\frac{3}{5} & 0 \\
 \bar{A} & \overline{\Yasymm} & 1 & 1 & 0 & -5 & \frac{3}{5} & 0 \\
 Q & \Yfund & \Yfund & 1 & 1 & 3 & \frac{1}{5} & \frac{3}{5} \\
 \bar{Q} & \overline{\Yfund} & 1 & \Yfund & -1 & 3 &
   -\frac{1}{5} & \frac{3}{5} \\ \hline \hline
 & SU(6) & SU(5)_Q & SU(5)_{\bar{Q}} & U(1)_1 & U(1)_2 & U(1)_3 & U(1)_R
  \\ \hline
 a & \Ythree & 1 & 1 & 0 & -5 & 0 & 0 \\
 q & \Yfund & \overline{\Yfund} & 1 & \frac{2}{3} & 2 & \frac{1}{3} &
\frac{2}{5} \\
 \bar{q} & \overline{\Yfund} & 1 & \overline{\Yfund} & -\frac{2}{3} & 2  &
- -\frac{1}{3}
    & \frac{2}{5} \\
 q_6 & \Yfund & 1 & 1 & \frac{5}{3} & 5 & -\frac{2}{3} & 1 \\
 \bar{q}_6 & \overline{\Yfund} & 1 & 1 & -\frac{5}{3} & 5 & \frac{2}{3} & 1  
\\
M_0& 1 & \Yfund & \Yfund & 0 & 6 & 0 & \frac{6}{5} \\
M_2& 1 & \Yfund & \Yfund & 0 & -4 & 0 & \frac{6}{5} \\
(M_2)_{6i} & 1 & 1 & \Yfund & -1 & -7 & 1 & \frac{3}{5} \\
(M_2)_{i6} & 1 & \Yfund & 1 & 1 & -7 & -1 &\frac{3}{5} \\
(M_2)_{66} & 1 & 1 & 1 & 0 & -10 & 0 & 0 \\
\end{array}
\end{equation}
with a superpotential in the magnetic $SU(6)$ theory
\begin{equation}
\label{SU5with5}
W=M_0 q a^2 \bar{q} + M_2 q \bar{q} + (M_2)_{6i} q_6 \bar{q}_i +
(M_2)_{i6} q_i \bar{q}_6 + (M_2)_{66} q_6 \bar{q}_6 + q_6 a^2 \bar{q}_6.
\end{equation}

Next, we construct a dual description for $SU(5)$ with $F=4$.
This dual can be obtained in two different
ways: either by higgsing the $SU(6)$ theory with  $\Ythree + 5(\Yfund
+\overline{\Yfund})$ or by integrating out a flavor from $SU(5)$ with
$F=5$. Here, we consider the latter possibility. Adding a mass term
$m q_5 \bar{q}_5$ corresponds to adding a term $m (M_0)_{55}$ to the
superpotential of equation~\ref{SU5with5}. This gives VEVs to $q_5$,
$\bar{q}_5$, and $a$ in analogy to the case of integrating out a flavor
from $SU(6)$ with $\Ythree + 6(\Yfund
+\overline{\Yfund})$. These
VEVs break $SU(6)$ to $Sp(4)$ and the resulting dual pair is given
in the table below.
\begin{equation}
\label{eq:su5four}
\begin{array}{c|ccccccc}
 & SU(5) & SU(4)_Q & SU(4)_{\bar{Q}} & U(1)_1 & U(1)_2 & U(1)_3 & U(1)_R
 \\ \hline
A & \Yasymm & 1 & 1 & 1 & 0 & -4 & 0 \\
\bar{A} & \overline{\Yasymm} & 1 & 1 & -1 & 0 & -4 & 0 \\
Q & \Yfund & \Yfund & 1 & 0 & 1 & 3 & \frac{1}{2} \\
\bar{Q} & \overline{\Yfund} & 1 & \Yfund & 0 & -1  & 3 & \frac{1}{2}
  \\ \hline \hline
 & Sp(4) & SU(4)_Q & SU(4)_{\bar{Q}} & U(1)_1 & U(1)_2 & U(1)_3 & U(1)_R
  \\ \hline
 a & \Yasymm & 1 & 1 & 0 & 0 & -8 & 0 \\
 q & \Yfund & \overline{\Yfund} & 1 & -\frac{1}{2} & 1 & 1 & \frac{1}{2} \\
 \bar{q} & \overline{\Yfund} & 1 & \overline{\Yfund} & \frac{1}{2} & -1 & 1
&
   \frac{1}{2} \\
 q_6 & \Yfund & 1 & 1 & \frac{3}{2} & 2 & 4 & 1 \\
 \bar{q}_6 & \overline{\Yfund} & 1 & 1 & -\frac{3}{2} & -2 & 4 & 1 \\
M_0& 1 & \Yfund & \Yfund & 0 & 0 & 6 & 1 \\
M_2& 1 & \Yfund & \Yfund & 0 & 0 & -2 & 1 \\
(M_2)_{6i} & 1 & 1 & \Yfund & -2 & -1 & -5 & \frac{1}{2} \\
(M_2)_{i6} & 1 & \Yfund & 1 & 2 & 1 & -5 &\frac{1}{2} \\
(M_2)_{66} & 1 & 1 & 1 & 0 & 0 & -8 & 0 \\
\end{array}
\end{equation}
The superpotential for the magnetic $Sp(4)$ theory is
\begin{equation}
\label{SU5with4}
W=M_0 q a \bar{q} + M_2 q \bar{q} + (M_2)_{6i} q_6 \bar{q}_i +
(M_2)_{i6} q_i \bar{q}_6 + (M_2)_{66} q_6 \bar{q}_6 + q_6 a \bar{q}_6.
\end{equation}

Integrating out one more flavor in the electric $SU(5)$ theory with
$F=4$ completely
breaks the gauge group of the dual theory. This is in complete analogy
to integrating out one flavor from the $SU(6)$ theory with $\Ythree +  
5(\Yfund
+\overline{\Yfund})$: one obtains the correct confining spectrum and
several terms in the confining superpotential.

There are several other non-trivial dualities which can be derived from
these duals by either higgsing the gauge group or adding mass terms
for quark antiquark pairs. Some of these duals are summarized in
Fig.~\ref{tab:nice}.
The figure elucidates one of the main lesson of this section:
theories with the same gauge group and gauge degrees of freedom
can have very different duals, depending on the ``meson" content and
the tree-level superpotential. For example, $SU(6)$ with
$\Ythree + 6(\Yfund + \overline{\Yfund})$ is self-dual, it is also
dual to $SU(5)$ with $\Yasymm + \overline{\Yasymm} + 5 (\Yfund +
\overline{\Yfund})$ and dual to $SU(4)$ with $2 \, \Yasymm + 5 (\Yfund
+\overline{\Yfund})$, depending on the ``meson" content.

\begin{figure}

\begin{picture}(462,250)(-60,150)
\thinlines

\put(40,368){\vector(0,-1){20}}
\put(40,308){\vector(0,-1){20}}
\put(40,248){\vector(0,-1){20}}

\put(120,308){\vector(0,-1){20}}
\put(120,248){\vector(0,-1){20}}

\put(200,308){\vector(0,-1){20}}
\put(200,248){\vector(0,-1){20}}

\put(280,308){\vector(0,-1){20}}
\put(280,248){\vector(0,-1){20}}

\put(73,367){\vector(1,-1){15}}
\put(73,307){\vector(1,-1){15}}
\put(73,247){\vector(1,-1){15}}

\put(152,310){\vector(1,0){20}}
\put(232,310){\vector(1,0){20}}
\put(152,250){\vector(1,0){20}}
\put(232,250){\vector(1,0){20}}
\put(152,190){\vector(1,0){20}}
\put(232,190){\vector(1,0){20}}

\thicklines

\put(-60,180){\framebox(380,300)}

\put(-4,180){\line(0,5){300}}
\put(0,180){\line(0,5){300}}
\put(80,180){\line(0,5){300}}
\put(160,180){\line(0,5){300}}
\put(240,180){\line(0,5){300}}

\put(-60,418){\line(5,0){380}}
\put(-60,422){\line(5,0){380}}
\put(-60,360){\line(5,0){380}}
\put(-60,300){\line(5,0){380}}
\put(-60,240){\line(5,0){380}}
\put(-60,180){\line(5,0){380}}

\put(-45,385){\bf F=6}
\put(-45,325){\bf F=5}
\put(-45,265){\bf F=4}
\put(-45,205){\bf F=3}

\put(10,445){\bf SU(6) $ \Ythree $}
\put(90,445){\bf SU(5) $ \Yasymm \ \overline{\Yasymm} $}
\put(170,445){\bf SU(4) $ \Yasymm \ \Yasymm $}
\put(250,445){\bf Sp(4) $ \Yasymm $}

\put(10,385){$ SU(6) \ \Ythree $}
\put(100,385){ -- }
\put(180,385){ -- }
\put(260,385){ -- }

\put(10,325){$ Sp(4) \ \Yasymm $}
\put(90,325){$ SU(6) \ \Ythree $}
\put(170,325){$ SU(6) \ \Ythree $}
\put(250,325){$ SU(6) \ \Ythree $}

\put(10,265){ confining }
\put(90,265){$ Sp(4) \ \Yasymm $}
\put(170,265){$ Sp(4) \ \Yasymm $}
\put(250,265){$ Sp(4) \ \Yasymm $}

\put(10,218){ confining, }
\put(20,208){ chiral }
\put(10,198){ symmetry }
\put(13,188){ breaking }

\put(90,205){ confining }
\put(170,205){ confining }
\put(250,205){ confining }

\end{picture}
\caption{\label{tab:nice}
The chain of theories obtained from the self-dual of an $SU(6)$
theory with the three-index antisymmetric tensor and six flavors.
The gauge group and the tensor field content of the electric
theories are indicated in the top row. The first column gives the
number of flavors. The confining theories are identified as such in
the table, for the others we give the dual gauge group and tensor
field content. The arrows depict possible flows between these theories
either by higgsing the gauge group or by adding mass terms.}
\end{figure}

We see in these examples that magnetic theories with very
similar ``meson" fields can be dual to electric theories with radically
different
gauge groups and/or matter content.
This suggests that finding the field content of a dual for a theory
with tensors is quite difficult in general.
The potential presence of accidental symmetries
complicates finding duals even more, since one of the main tools for
identifying dualities is to require that global symmetries and their
anomalies match in the electric and magnetic theories.
When two theories have differing ultraviolet symmetries, it is difficult
to decide whether there is a duality connecting the two theories,
with some symmetries being accidental,
or whether there is no duality at all.

\subsection{Multiple self-duals}

In this section we present another example of multiple
dualities, similar to the $SU(2)$ theory presented in Section 2.
We consider $SU(4)$ with $2\, \Yasymm +4(\Yfund + \overline{\Yfund} )$.
An $Sp(4)$ dual for this theory can be obtained by
higgsing and integrating
out a flavor from the self-dual of $SU(6)$ with $\Ythree$.
This $SU(4)$ theory has however several self-dual descriptions
as well. These are described in Table~\ref{multipletable}.

\begin{table}
\[
\begin{array}{c|ccccccc}
& SU(4) & SU(2) & SU(4) & SU(4) & U(1)_1 & U(1)_2 & U(1)_R \\
\hline
A & \Yasymm & \Yfund & 1 & 1 & 0 & 2 & 0 \\
Q & \Yfund & 1 & \Yfund & 1 & 1 & -1 & \frac{1}{2} \\
\bar{Q} & \overline{\Yfund} & 1 & 1 & \Yfund & -1 & -1 & \frac{1}{2} \\  
\hline
\hline
& SU(4)_{D_1} & SU(2) & SU(4) & SU(4) & U(1)_1 & U(1)_2 &
U(1)_R \\ \hline
a & \Yasymm & \Yfund & 1 & 1 & 0 & 2 & 0 \\
q & \Yfund & 1 & \overline{\Yfund} & 1 & -1 & -1 & \frac{1}{2} \\
\bar{q} & \overline{\Yfund} & 1 & 1 & \overline{\Yfund} & 1 &
- -1 & \frac{1}{2} \\
M_0 & 1 & 1 & \Yfund & \Yfund & 0 & -2 & 1 \\
M_2 & 1 & 1 & \Yfund & \Yfund & 0 & 2 & 1 \\
B & 1 & \Yfund & \Yasymm & 1 & 2 & 0 & 1 \\
\bar{B} & 1 & \Yfund & 1 & \Yasymm & -2 & 0 & 1 \\ \hline \hline
& SU(4)_{D_2} & SU(2) & SU(4) & SU(4) & U(1)_1 & U(1)_2 &
U(1)_R \\ \hline
a & \Yasymm & \Yfund & 1 & 1 & 0 & 2 & 0 \\
q & \Yfund & 1 & \overline{\Yfund} & 1 & 1 & -1 & \frac{1}{2} \\
\bar{q} & \overline{\Yfund} & 1 & 1 & \overline{\Yfund} & -1 &
- -1 & \frac{1}{2} \\
M_0 & 1 & 1 & \Yfund & \Yfund & 0 & -2 & 1 \\
M_2 & 1 & 1 & \Yfund & \Yfund & 0 & 2 & 1 \\ \hline \hline
& SU(4)_{D_3} & SU(2) & SU(4) & SU(4) & U(1)_1 & U(1)_2 &
U(1)_R \\ \hline
a & \Yasymm & \Yfund & 1 & 1 & 0 & 2 & 0 \\
q & \Yfund & 1 & \Yfund & 1 & -1 & -1 & \frac{1}{2} \\
\bar{q} & \overline{\Yfund} & 1 & 1 & \Yfund & 1 &
- -1 & \frac{1}{2} \\
B & 1 & \Yfund & \Yasymm & 1 & 2 & 0 & 1 \\
\bar{B} & 1 & \Yfund & 1 & \Yasymm & -2 & 0 & 1
\end{array}
\]
\caption{\label{multipletable}The multiple self-duals of $SU(4)$ with
$2\, \protect\Yasymm +4(\protect\Yfund +\overline{\protect\Yfund})$.}
\end{table}
Note that the representations and charges of the dual quarks differ in the
three duals. The 't~Hooft anomaly matching and the operator map
is again straightforward if one includes the appropriate superpotentials
for the different duals:

\begin{eqnarray} W_{D_1} &=& M_0 qa^2\bar{q} + M_2 q\bar{q} + B aq^2
+\bar{B}a\bar{q}^2 \nonumber \\
W_{D_2} &=& M_0 qa^2\bar{q} + M_2 q\bar{q} \\
W_{D_3} &=&  B aq^2 +\bar{B}a\bar{q}^2 \nonumber
\end{eqnarray}

As a consistency check on this duality we consider giving giving a VEV to
one of
the antisymmetric tensors. This breaks both the electric and the magnetic
$SU(4)$ gauge group to $Sp(4)$, leaving one antisymmetric tensor and
eight flavors of $Sp(4)$. A self-dual of this $Sp(4)$ theory has been
described in Ref.~\cite{oursp}. Our $D_1$ self-dual exactly reproduces
the dual of \cite{oursp}. Our other two duals lead to
new dualities for the $Sp(4)$ theory. Note that in the latter case
the fundamental fields of the dual $Sp(4)$ theory do not combine
into complete representations of the global $SU(8)$ group,
only an $SU(4)\times SU(4) \times U(1)$ subgroup is manifest in the dual.
The full $SU(8)$ global symmetry arises as an accidental symmetry of
the infrared.

As a consistency check, one can further higgs the $Sp(4)$ gauge group
to $SU(2)\times SU(2)$,
by giving a VEV to the to the remaining antisymmetric tensor. One finds
that the three duals flow to two copies of the three self-duals
of $SU(2)$ with 8 doublets described in Section 2, providing another
consistency check on these dualities.

\subsection{$SU(2N)$ with $\protect\Yasymm +\overline{\protect\Yasymm} +
4(\protect\Yfund +\overline{\protect\Yfund})$}

In this section we describe the generalization of one of the $SU(4)$  
self-duals
presented in the previous section to $SU(2N)$ with
$\Yasymm +\overline{\Yasymm} +4(\Yfund +\overline{\protect\Yfund})$. The
pair of dual theories is given in the table below.
\begin{equation}
\label{su2n}
\begin{array}{c|ccccccc}
 &  SU(2N) & SU(4)_Q & SU(4)_{\bar{Q}} & U(1)_1 & U(1)_2 & U(1)_3 & U(1)_R
 \\ \hline
A & \Yasymm & 1 & 1 & 1 & 0 & -4 & 0 \\
\bar{A} & \overline{\Yasymm} & 1 & 1 & -1 & 0 & -4 & 0 \\
Q & \Yfund & \Yfund & 1 & 0 & 1 & 2N-2 & \frac{1}{2} \\
\bar{Q} & \overline{\Yfund} & 1 & \Yfund & 0 & -1  & 2N-2 & \frac{1}{2}\\
\hline \hline
 &  SU(2N)_D & SU(4)_Q & SU(4)_{\bar{Q}} & U(1)_1 & U(1)_2 & U(1)_3 & U(1)_R
 \\ \hline
a & \Yasymm & 1 & 1 & 1 & 0 & -4 & 0 \\
\bar{a} & \overline{\Yasymm} & 1 & 1 & -1 & 0 & -4 & 0 \\
q & \Yfund & \overline{\Yfund} & 1 & 0 & 1 & 2N-2 & \frac{1}{2} \\
\bar{q} & \overline{\Yfund} & 1 & \overline{\Yfund} & 0 & -1  & 2N-2 &
\frac{1}{2}\\
M_k & 1 & \Yfund & \Yfund & 0&0&4N-4-8k&1\\
H_m & 1 & \Yasymm & 1 & -1 & 2 & 4N-8-8m & 1 \\
\bar{H}_m & 1& 1 & \Yasymm & 1 & -2 & 4N-8-8m & 1
\end{array}
\end{equation}
where $k=0,\ldots,N-1$, $m=0,\ldots,N-2$, and
with the superpotential in the dual magnetic theory given by
\begin{equation}
\label{su(2n)}
W=\sum_k M_k q \bar{q} (a\bar{a})^{N-1-k} + \sum_m \Big( H_m q^2 a
(a\bar{a})^{N-2-m}+\bar{H}_m \bar{q}^2 \bar{a} (a\bar{a})^{N-2-m} \Big)
\end{equation}

The 't~Hooft anomaly matching conditions are again straightforward to check,
while the mapping of flat directions is given by

\[
\begin{array}{cc}
Q\bar{Q}(A\bar{A})^k \leftrightarrow M_k \quad &
A^{N-1} Q^2 \leftrightarrow a^{N-1} q^2 \\
Q^2 \bar{A} (A\bar{A})^k \leftrightarrow H_k \quad & \bar{A}^{N-1}
\bar{Q}^2 \leftrightarrow \bar{a}^{N-1} \bar{q}^2 \\
\bar{Q}^2 {A} (A\bar{A})^k \leftrightarrow \bar{H}_k \quad & (A\bar{A})^k
\leftrightarrow (a\bar{a})^k \\
\qquad A^N \leftrightarrow a^N & A^{N-2} Q^4 \leftrightarrow a^{N-2} q^4 \\
\qquad \bar{A}^N \leftrightarrow \bar{a}^N & \bar{A}^{N-2} \bar{Q}^4
\leftrightarrow \bar{a}^{N-2} \bar{q}^4 \end{array}
\]

Note that a similarly constructed candidate for a self-dual
to $SU(2N+1)$ with $\Yasymm +\overline{\Yasymm}
+4(\Yfund +\overline{\protect\Yfund})$
fails to match the global anomalies.

As a consistency check on the above duality for $SU(2N)$ we consider
breaking $SU(2N)$ to $SU(2)^N$ by giving a VEV to the fields $A$ and
$\bar{A}$. The result is $N$ copies of the $SU(2)$ self-dual with
$8$ doublets presented in Section 2. Another check is to integrate out
one flavor from the electric theory. On the magnetic side this corresponds
to completely higgsing the gauge group since the operator
$q(a\bar{a})^{N-1}\bar{q}$ is forced to have an expectation value by
the $M_0$ equation of motion. The massless fields exactly correspond to the
confining spectrum of $SU(2N)$ with
$\Yasymm +\overline{\Yasymm} +3(\Yfund +\overline{\protect\Yfund})$
described in Ref.~\cite{s-confinement}. Again, part of the
confining superpotential is reproduced by the tree-level
superpotential of equation~\ref{su(2n)} while the rest is presumably  
generated by instanton effects.

While there is no self-dual for $SU(2N+1)$ with $\Yasymm +\overline{\Yasymm}
+4(\Yfund +\overline{\protect\Yfund})$, one can derive a non-trivial
dual for this theory by breaking the $SU(2N+2)$ gauge group of this  
self-dual to $SU(2N+1)$ with an expectation value for one flavor of quarks.
The electric theory becomes $SU(2N+1)$ with $\Yasymm +\overline{\Yasymm}
+4(\Yfund +\overline{\protect\Yfund})$, while the dual magnetic
theory is still $SU(2N+2)$ with $\Yasymm +\overline{\Yasymm}
+4(\Yfund +\overline{\protect\Yfund})$, but with a different combination
of singlets and tree-level superpotential. Just like
in the derived dualities of Section 3.2, the $SU(4)\times SU(4)$
global symmetry is not explicit in the magnetic theory, but only restored
in the infrared.

%%%%%%%%%%%%%%%%%%%%%%%%%%%%%%%%%
\subsection{$SO(N)$ with spinors and $N-4$ vectors}

In this section, we present a series of self-dual theories, all of which have
multiple self-duals. The theories we examine have $SO(N)$ gauge groups
with $N-4$ vectors as well as some spinor representations.
The duals of these theories contain the same gauge degrees of freedom
as the electric theory, and some additional gauge  singlets.
The gauge singlet ``meson" fields correspond to composite operators of
the electric theory made up of two spinors and varying numbers of vectors.
In the simplest self-dual, the $SO(N)$ vectors remain fundamentals of
the $SU(N-4)$ global symmetry, while the dual spinors
become antifundamentals under their global symmetry.
Thus, in some sense only the spinors are being ``dualized" in these duals.

As an explicit example we describe an $SO(8)$ theory with 4 spinors and
4 vectors in detail. This theory is particularly interesting because
$SO(8)$ has a group automorphism.
With this choice of matter content the automorphism implies
a $Z_2$ symmetry which exchanges
spinors and vectors. The duality is as follows:

\begin{equation} \begin{array}{c|c|cccc}
& SO(8) & SU(4) & SU(4) & U(1) & U(1)_R \\ \hline
S & 8_s & \Yfund & 1 & 1 & \frac{1}{4} \\
V & 8_v & 1 & \Yfund & -1 & \frac{1}{4} \\ \hline \hline
& SO(8)_{D1} & SU(4) & SU(4) & U(1) & U(1)_R \\ \hline
s & 8_s & \overline{\Yfund} & 1 & 1 & \frac{1}{4} \\
v & 8_v & 1 & \Yfund & -1 & \frac{1}{4} \\
M_0 & 1 & \Ysymm & 1 & 2 & \frac{1}{2} \\
M_2 & 1 & \Yasymm & \Yasymm & 0 & 1 \\
M_4 & 1 & \Ysymm & 1 & -2 & \frac{3}{2} \end{array}
\end{equation}
with a dual  superpotential:
\begin{equation} W_{\rm magn}=M_0 s^2v^4+M_2 s^2v^2+M_4 s^2.
\end{equation}
Note that the ``meson" content of this dual does not have a manifest $Z_2$
symmetry, thus the exchange symmetry is an accidental infrared symmetry
in the dual description, in analogy to the symmetries of the $SU(2)$ theory
discussed in Section 2.  Again, there are a number of other self-duals,
which are summarized in Table~\ref{so8table}.

\begin{table}
\[
\begin{array}{c|c|cccc}
& SO(8)_{D2} & SU(4) & SU(4) & U(1) & U(1)_R \\ \hline
s & 8_s & \overline{\Yfund} & 1 & 1 & \frac{1}{4} \\
v & 8_v & 1 & \Yfund & -1 & \frac{1}{4} \\
M_0 & 1 & \Ysymm & 1 & 2 & \frac{1}{2} \\
M_4 & 1 & \Ysymm & 1 & -2 & \frac{3}{2} \\
\hline \hline
& SO(8)_{D3} & SU(4) & SU(4) & U(1) & U(1)_R \\ \hline
s & 8_s & \Yfund & 1 & 1 & \frac{1}{4} \\
v & 8_v & 1 & \Yfund & -1 & \frac{1}{4} \\
M_2 & 1 & \Yasymm & \Yasymm & 0 & 1 \\
\hline \hline
& SO(8)_{D4} & SU(4) & SU(4) & U(1) & U(1)_R \\ \hline
s & 8_s & \Yfund & 1 & 1 & \frac{1}{4} \\
v & 8_v & 1 & \overline{\Yfund} & -1 & \frac{1}{4} \\
\bar{M_0} & 1 &  1 & \Ysymm & -2 & \frac{1}{2} \\
M_2 & 1 & \Yasymm & \Yasymm & 0 & 1 \\
\bar{M_4} & 1 & 1 & \Ysymm & 2 & \frac{3}{2} \\
\hline \hline
& SO(8)_{D5} & SU(4) & SU(4) & U(1) & U(1)_R \\ \hline
s & 8_s & \overline{\Yfund} & 1 & 1 & \frac{1}{4} \\
v & 8_v & 1 & \overline{\Yfund} & -1 & \frac{1}{4} \\
M_0 & 1 & \Ysymm & 1 & 2 & \frac{1}{2} \\
\bar{M_0} & 1 & 1 & \Ysymm & -2 & \frac{1}{2} \\
M_2 & 1 & \Yasymm & \Yasymm & 0 & 1 \\
M_4 & 1 & \Ysymm & 1 & -2 & \frac{3}{2} \\
\bar{M_4} & 1 & 1 & \Ysymm & 2 & \frac{3}{2}
\end{array}
\]
\caption{\label{so8table} Multiple self-duals of $SO(8)$ with 4 spinors
and 4 vectors. }
\end{table}
 The operator mapping is:
\begin{equation}
\begin{array}{cc}
S^2 \leftrightarrow M_0  \\
V^2 \leftrightarrow \bar{M_0}  \\
S^2V^2 \leftrightarrow M_2   \\
S^2V^4 \leftrightarrow M_4    \\
S^4V^2 \leftrightarrow \bar{M_4}
\end{array}
\end{equation}

It is interesting to examine the possibility that
this theory is in an interacting non-Abelian Coulomb phase, i.e. the gauge
coupling runs to a non-trivial fixed point in the infrared.
Then, due to  the $Z_2$ symmetry, the anomalous dimensions
of the spinor and vector quarks must be the same. This
uniquely identifies
the superconformal $R$ symmetry (which is related to the scaling
dimensions~\cite{lectnotes}) as the $R$ symmetry whose
charges we indicated in the tables above.
Recalling the relation that the scaling dimension is $\frac{3}{2}
R_{\rm sc}$ for fields in the superconformal algebra and the fact
that gauge-invariant chiral superfields cannot have a
dimension less than one (chiral superfields with $R_{\rm sc}$
less than ${{2}\over{3}}$ are necessarily free-fields and decouple from
the superconformal theory)
we find  that the chiral superfields  $M_0$ and $\bar{M_0}$
are free in the infrared, while the
remaining degrees of freedom are interacting with scaling dimensions
given by $\frac{3}{2} R_{\rm sc}$. Thus this
theory may provide a realization of the exotic phenomena
suggested in reference~\cite{LST}.

Similar self-duals can be found for numerous other $SO(N)$ theories with
spinors and $N-4$ vectors. The properties of these duals are summarized in
Table~\ref{SOtable}.
We use the following notation for the matter content:
$(N_S,N_C,N_V)$,
which denotes $N_S$ spinors, $N_C$ conjugate spinors, and $N_V$ vectors.
For $SO(N)$ groups with $N$ odd only two numbers are given, the first for
spinors, the second for vectors.
The matter content of the self-dual theories is determined by
assigning the vectors $R$-charge 0, and adding as many spinors as required
for the spinors to have $R$-charge $\frac{1}{2}$.
\begin{table}
\begin{displaymath}
\begin{array}{c|c|l}
{\rm group} & {\rm content}     &       {\rm ``mesons"} \\ \hline \hline
SO(4) & (8,8,0)  & \sim SU(2)\times SU(2) \; \mbox{with}\;
8(\Yfund ,1) + 8 (1,\Yfund) \\
SO(5) & (8,1) & \sim Sp(4) \; \mbox{with}\;  8\, \Yfund + \Yasymm \\
SO(6) & (4,4,2)  & \sim SU(4) \; \mbox{with}\;
4(\Yfund +\overline{\Yfund}) + 2\, \Yasymm \\
SO(7)   & (4,3)         &       s^2,s^2v,s^2v^2,s^2v^3  \\
SO(8)   & (4,0,4)       &       s^2,s^2v^2,s^2v^4 \\
SO(8)   & (2,2,4)  &  s^2,s^2v^2,s^2v^4,c^2,c^2v^2,c^2v^4,scv,scv^3 \\
SO(8)   & (3,1,4) & s^2, s^2v^2,s^2v^4,c^2,c^2v^4,scv,scv^3 \\
SO(9)   & (2,5) &               s^2,s^2v,s^2v^2,s^2v^3,s^2v^4,s^2v^5 \\
SO(10)  & (2,0,6)       &       s^2v,s^2v^3,s^2v^5 \\
SO(10)  & (1,1,6)  & s^2v,s^2v^5,c^2v,c^2v^5,sc,scv^2,scv^4,scv^6 \\
SO(11)  & (1,7) &               s^2v,s^2v^2,s^2v^5,s^2v^6 \\
SO(12)  & (1,0,8)       &       s^2v^2,s^2v^6 \\
\end{array}
\end{displaymath}
\caption{\label{SOtable}
The  series of self-dual $SO(N)$ theories with $N-4$ vectors. The
first column gives the gauge group, the second the matter content of the
electric theory, while the third column gives the additional gauge singlet
``mesons" one needs to include into the dual magnetic theory (with the
appropriate tree-level superpotential). All of these theories turn out to
have multiple dualities. As noted, the first three examples are equivalent
to theories discussed in previous sections.}
\end{table}

Note that these self-duals flow to each other when giving a VEV to one
vector. Both the electric and magnetic gauge groups break to $SO(N-1)$.
One vector is eaten, while $N-5$ vectors remain in the theory. It is
straightforward to check, that the resulting theory is exactly the
corresponding self-dual of $SO(N-1)$ in our table.

As a consistency check, one can give VEVs to all $SO(N)$
vectors, breaking $SO(N)$ to $SO(4)\sim SU(2)\times SU(2)$.
It is easy to check that all self-duals of Table~\ref{SOtable}
reduce to two copies of one of the three $SU(2)$ self-duals
discussed in Section 2.

Another check on these dualities consist of integrating out
vectors from the theory. After integrating out one
vector the electric theory confines without chiral
symmetry breaking and with a confining superpotential
(s-confines)~\cite{s-confinement}. The magnetic theory
also confines, and it is straightforward to check that
the confined  degrees of freedom are correctly reproduced.
However, just like in the duality of~\cite{oursp}, the
origin of some of the terms in the confining superpotential on the magnetic
side is unknown, but may be due to instantons.

Finally, we consider giving masses to spinors in those
theories where the spinors are in real representations.
When integrating out all spinors the electric theory becomes
$SO(N)$ with $N-4$ vectors, which has two branches of vacua~\cite{IntSeib}.
One branch has a dynamically generated superpotential while the other branch
is confining with no superpotential. To check how this arises in the
magnetic theories we consider the example
of $SO(6)$ with $(4,4,2)$ which is equivalent to
$SU(4)$ with $2\, \Yasymm + 4(\Yfund + \overline{\Yfund})$.
Integrating out the spinors in $SO(6)$
amounts to integrating out the $SU(4)$ flavors, which we
do one at a time. Adding a mass term for one flavor gives
non-zero VEVs to fields in the magnetic theory by the
``meson" equation of motion which breaks the magnetic $SU(4)$
gauge group completely. After identifying the uneaten light
degrees of freedom we obtain the confined spectrum and part of the
confining superpotential
of $SU(4)$ with $2\, \Yasymm + 3(\Yfund +\overline{\Yfund})$,
described in Ref.~\cite{s-confinement}. The remaining part of the
confining superpotential is presumably generated by instanton
effects in the completely broken $SU(4)$ gauge group.
Thus after integrating out one flavor the magnetic theory
reproduces the confining theory with 3 flavors.
It has been shown in Ref.~\cite{s-confinement}, that integrating
out the remaining three flavors does result in the correct
description of the theory in both branches of vacua.

\section{Conclusions}
We have presented a large set of new self-dual $N=1$ SUSY gauge theories.
Starting from these self-duals one can obtain many new nontrivial
dualities by integrating out flavors or by higgsing the gauge group.
In addition to the derived dualities discussed here, a large class of
duals for $SU(N)$ theories with antisymmetric tensors can be derived
by giving expectation values to the spinor representations of our
$SO(N)$ self-duals.
A common feature of many of these derived dualities is that they have some
accidental global symmetries, which emerge only in the infrared and
are not explicit in the ultraviolet description. These accidental symmetries
can make finding new dualities very difficult. Usually, one is looking for
dual pairs by requiring that the global symmetries and their anomalies
in the two theories match. When some of the global symmetries are accidental
symmetries, present only in the infrared, we loose our most powerful
tool for identifying dualities.

Another interesting feature of the presented dualities is that they do not
seem to follow any obvious pattern. For example, $SU(5)$ with $2 \, \Yasymm
+ 5 (\Yfund + \overline{\Yfund})$ is dual to $SU(6)$ with $\Ythree + 6
(\Yfund + \overline{\Yfund})$, while $SU(5)$ with $2 \, \Yasymm +
4 (\Yfund + \overline{\Yfund})$ is dual to $Sp(4)$ with $\Yasymm + 10\,
\Yfund$. Just by changing the number of flavors, the gauge group and the
matter content of the dual theory changes completely. Furthermore, theories
with identical gauge groups and gauge degrees of freedom but with
different ``meson" content and superpotentials can be dual to very
different theories. For example,  $SU(6)$ with $\Ythree + 6
(\Yfund + \overline{\Yfund})$ is self-dual, dual to $SU(5)$ with
$2 \, \Yasymm + 5 (\Yfund + \overline{\Yfund})$ or dual to $SU(4)$ with
$2 \, \Yasymm + 5 (\Yfund + \overline{\Yfund})$, depending on the gauge
singlet ``meson" content and the tree-level superpotential.

These features of duality suggest that developing a systematic approach to
finding $N=1$ duals for theories with arbitrary matter content might be
a very difficult task, which will likely require more insight into the
dynamics of supersymmetric gauge theories than is presently available.

\section*{Acknowledgments}
We are grateful to Lisa Randall and Cumrun Vafa for useful discussions.
C.C. and W.S. are supported in part by the U.S.
Department of Energy under cooperative
agreement \#DE-FC02-94ER40818. M.S. is supported by the U.S.
Department of Energy under grant \#DE-FG02-91ER40676.
J.T. is supported by the National Science Foundation under grant
PHY-95-14797.

\end{document}